# Towards SAR Tomographic Inversion via Sparse Bayesian Learning

Kun Qian[a], Yuanyuan Wang[a,b], Xiaoxiang Zhu[a,b]
[a] Technical University of Munich, Munich, Germany
[b] German Aerospace Center, Weßlinng, Germany

## Abstract

SAR tomographic inversion (TomoSAR) has been widely employed for 3-D urban mapping. TomoSAR is essentially a spectral estimation problem. Existing algorithms are mostly based on an explicit inversion of the SAR imaging model, which are often computationally expensive for large scale processing. This is especially true for compressive sensing based TomoSAR algorithms. Previous literature showed perspective of using data-driven methods like PCA and kernel PCA to decompose the signal and reduce the computational complexity of parameter inversion. This paper gives a preliminary demonstration of a new data-driven TomoSAR method based on *sparse Bayesian learning*. Experiments on simulated data show that the proposed method significantly outperforms the previously proposed PCA and KPCA methods in estimating the steering vectors of the scatterers. This gives us a perspective of using data-drive approach or combining data-driven and model-driven approach for high precision tomographic inversion for large areas.

## 1 Introduction

SAR tomography (TomoSAR) is widely used for the reconstruction of 3-D urban models. It extends the synthetic aperture principle into the elevation direction and allows an estimation of scatterers' 3-D distribution. Usually, TomoSAR is achieved via an explicit inversion of the SAR imaging model, which is computationally expensive for large area processing, since it requires the inversion of a very large model matrix and the detection of local maxima in the retrieved reflectivity profile.

In order to circumvent such high computational cost, data-driven methods can be employed. In recent years, [1] exploited the potential of principle component analysis (PCA) in TomoSAR inversion to blindly separate the phase of multiple scatters and resolve them. The principle of this PCA-based method is to approximated the steering vectors of individual scatters layovered in a single pixel by the dominant eigenvectors of the sample covariance matrix. However, such correspondence between eigenvectors and steering vectors is not fulfilled in theory, since the steering vectors of different scatterers are seldomly mutually orthogonal. Therefore, [2] introduced a kernel PCA (KPCA) method to transfer the data into high dimension thus relax the requirement of signal orthogonality.

So far, none of the abovementioned data-driven methods addressed the scatterers separation in the super resolution regime. In order to study the potential of more advanced data-driven methods like deep learning, as well as their perspective in super-resolving TomoSAR, this paper proposed a data-driven approach for TomoSAR inversion using sparse Bayesian learning (SBL) [3]. The TomoSAR inversion is formulated as basis selection in the proposed method. The steering matrix or the so-called mapping matrix, which comprises a span of steering vectors, is modelled as a redundant dictionary. The steering vectors of individual scatterers are selected from the dictionary via supervised learning.

## 2 TomoSAR Imaging Model and Problem Formulation

In this paper, we only consider the fully coherent multi-baseline InSAR model. Under such assumption, the canonical form of discrete SAR imaging model is given by

$$\mathbf{g} = \mathbf{R}\boldsymbol{\gamma}, \qquad (1)$$

where $\mathbf{g} \in \mathbb{C}^N$ is the multi-baseline SAR measurements vector, $\mathbf{R} \in \mathbb{C}^{N \times L}$ is the steering matrix, $\boldsymbol{\gamma} \in \mathbb{C}^L$ is the discrete reflectivity vector along the elevation direction, $\boldsymbol{\varepsilon} \in \mathbb{C}^N$ is the noise vector, and N, L are the number of acquisitions and the discretization level in the elevation direction, respectively. Generally, the discretization level is much larger than the number of SAR images, i.e. N<L.

In urban area, the reflectivity profile $\boldsymbol{\gamma}$ is usually consisted of the contribution from a few dominant scatterers, so that we can consider $\boldsymbol{\gamma}$ as a sparse signal in the object domain. In this vein, seeking reflectivity vector whose entries are predominantly zero while still allowing us to accurately approximate $\mathbf{g}$ is equivalent to representing $\mathbf{g}$ with a minimum number of steering vectors. This is a typical problem of basis selection. Specifically, it is desired to screen out the steering vectors of individual scatterers from an overcomplete steering matrix.

# 3 Related Work

## 3.1 PCA-based method

Linear PCA-based method [1] retrieves the steering vectors of individual scatterers by performing singular value decomposition (SVD) on the covariance matrix $\boldsymbol{C}$ of the signal:

$$\boldsymbol{C} = \boldsymbol{VDV^H} \qquad (2)$$

where $\mathbf{V}$ is the eigenvectors, and $\mathbf{D}$ is the diagonal matrix of eigenvalues. According to [1], the dominant eigenvectors correspond to steering vectors of individual scatterers. And hence, the steering vectors can be estimated by

$$\hat{\boldsymbol{r}}_i = \boldsymbol{v}_i \odot |\boldsymbol{v}_i|^{-1} \qquad (3)$$

where $\boldsymbol{v}_i$ is the columns of $\mathbf{V}$, and $\odot$ is the elementwise product. The steering vectors retrieved by this method are mutually orthogonal. Such orthogonality can never be fulfilled in reality. However, the orthogonality between two steering vectors increases, when the distance between the two scatterers increases or the number of acquisition increases. Therefore, PCA can also produce reasonable results in some circumstances.

## 3.2 Kernel PCA-based method

To circumvent the limitations of PCA-based methods, we proposed using KPCA in [2]. KPCA is a nonlinear extension of PCA. It maps the data to a much higher dimension before performing the PCA. The nonlinear transformation can be denoted as follows:

$$\Psi: \mathbb{C}^N \to F, \mathbf{g} \to \Psi(\mathbf{g}) \qquad (4)$$

where $F$ is the transformed vector space which may have arbitrary dimension. The KPCA of $\mathbf{C}$ is basically the eigenvalue decomposition of the covariance matrix in the transformed space which is

$$\boldsymbol{C_{\Psi\Psi}} = E[\Psi(\mathbf{g})\Psi(\mathbf{g})^H]. \qquad (5)$$

However, this nonlinear transformation and the higher dimensional covariance matrix are never explicitly evaluated. They are indirectly evaluated through the kernel trick.

Since the transformed signal can have infinite dimension, e.g. when using the Gaussian kernel, KPCA can greatly mitigate the non-orthogonality of the signals. Hence, it can provide superior results comparing to PCA, especially when the number of images is low, or when the two scatterers are similar in height or bright- ness. However, the performance of KPCA based methods is affected by the choice of the kernel as well as the hyperparameters of the kernel.

# 4 SAR Tomography via Sparse Bayesian Learning

As it is briefly mentioned previously, we have much less stacked SAR images than samples along elevation direction. In this context, TomoSAR inversion is actually an ill-posed inversion problem. Hence, a priori information is indispensable for constraining the solution. Sparsity of the reflectivity vector along the elevation direction is the most promising prior for TomoSAR inversion in urban area. State-of-the-art compressive sensing-based method, such as "SL1MMER" [4] uses the $L_1$ norm of the reflectivity vector as a sparsity-promoting function to guarantee a sufficiently sparse reconstruction. Unlike SL1MMER, which assume a predefined and fixed priori, SBL estimates parameterized priori from the data. [11] has extended the application of SBL to complex domain and [12][13] have validated the feasibility of using SBL for TomoSAR inversion. However, there is still a lack of a systematic study of SBL for TomoSAR inversion and the comparison to other data-driven methods as well as the state-of-the-art compressive sensing based methods.

Here we will firstly describe how to perform TomoSAR inversion via SBL. We assume the noise to be independent and and identically distributed (i.i.d.) complex circular Gaussian random variables. The likelihood of $\mathbf{g}$ hence follows the distribution of $\boldsymbol{\varepsilon}$, which is as follows

$$p(\boldsymbol{\varepsilon}) = \frac{1}{\pi^N \det(\mathbf{C}_{\varepsilon\varepsilon})} \exp(-\boldsymbol{\varepsilon}^H \mathbf{C}_{\varepsilon\varepsilon}^{-1} \boldsymbol{\varepsilon}). \qquad (6)$$

We can simply (6) using the residuals sum of squares (RSS) and obtain the following form

$$p(\mathbf{g}|\boldsymbol{\gamma},\sigma^2) = (\pi\sigma^2)^{-N} \exp\left(-\frac{1}{\sigma^2}\|\mathbf{g} - \mathbf{R}\boldsymbol{\gamma}\|_2^2\right), \qquad (7)$$

where $\sigma^2$ is the variance of the noise. An maximum a posteriori (MAP) maximizes the posterior showing in the equation (8)

$$p(\boldsymbol{\gamma}|\mathbf{g},\sigma^2) = \frac{p(\mathbf{g}|\boldsymbol{\gamma},\sigma^2)p(\boldsymbol{\gamma})}{p(\mathbf{g}|\sigma^2)}. \qquad (8)$$

The prior $p(\gamma)$ is chosen as $L_2$ norm in conventional Tikhonov regularization, and as $L_1$ norm in compressive sensing based methods. While no matter $L_2$ norm or $L_1$ norm is employed, the prior is predetermined and stay constant once it is given. In the proposed inversion method via SBL, we introduce a parameterized prior, which is governed by a set of hyperparameters that can be learned from data. This parameterized prior will be utilized as a Tikhonov regularizer in the following steps. We estimate each element of the parameterized prior in the learning procedure to form a sparse prior, thus promoting the sparsity of the estimated reflectivity profile. The hyperparameters represent the variance of the reflectivity vector along elevation direction at each discretization position. The parameterized prior is given by

$$p(\gamma | \mathbf{w}) = \prod_{i=1}^{L} (2\pi w_i)^{-\frac{1}{2}} \exp(-\frac{\gamma_i^2}{2w_i}). \quad (9)$$

where $\mathbf{w} = [w_1, w_2, ..., w_L]$ is a set of hyperparameters, and $\mathbf{C}_{\gamma\gamma} = diag(\mathbf{w})$ is the covariance matrix of the reflectivity vectors under the assumption that the scatterers are independent from each other. Under the SBL inference framework, we can iteratively update the hyperparameters and then achieve a sparse solution via pure Tikhonov regularization, thus taking the advantages of a sparse solution and computational efficiency. By substituting the parameterized prior $p(\gamma|\mathbf{w})$ into the equation (8), we can rewrite the posterior in the form of

$$p(\gamma | \mathbf{g}, \mathbf{w}, \sigma^2) = \frac{p(\mathbf{g} | \gamma, \sigma^2) p(\gamma | \mathbf{w})}{p(\mathbf{g} | \mathbf{w}, \sigma^2)}. \quad (11)$$

As it is described in [7], the optimization for estimating hyperparameter can be formed as type-II maximum likelihood or evidence maximization

$$\{\hat{\mathbf{w}}, \hat{\sigma}\} = \arg\max_{\mathbf{w}, \sigma} p(\mathbf{g} | \mathbf{w}, \sigma^2). \quad (12)$$

The marginal pdf $p(\mathbf{g}|\mathbf{w}, \sigma^2)$ is given by

$$p(\mathbf{g} | \mathbf{w}, \sigma^2) = \int p(\mathbf{g} | \gamma, \sigma^2) p(\gamma | w) d\gamma = (\pi)^{-N} |\mathbf{C}_{gg}| \exp(-\mathbf{g}^H \mathbf{C}_{gg}^{-1} \mathbf{g}), \quad (13)$$

where $|\cdot|$ denotes the determinant of a matrix and $\mathbf{C}_{gg} = \sigma^2 \mathbf{I} + \mathbf{R} \mathbf{C}_{\gamma\gamma} \mathbf{R}^H$. Taking the minus logarithms of equation (13) gives us efficient cost function of SBL

$$L = \ln\left(\pi^N \det(\mathbf{C}_{gg})\right) + \mathbf{g}^H \mathbf{C}_{gg}^{-1} \mathbf{g}, \quad (14)$$

and $\{\hat{\mathbf{w}}, \hat{\sigma}\} = \arg\min_{\mathbf{w}, \sigma} \{L\}$. According to [7], it is difficult to solve this minimization problem analytically. To cope with this minimization problem, MacKey fix-point approcach is usually employed, which is an iterative algorithm of updating the hyperparameters as follows

$$\hat{w}_i = \frac{\hat{\gamma}_i^2}{1 - w_i^{-1} \mathbf{C}_{\gamma\gamma}}, \quad (15)$$

$$\hat{\sigma}^2 = \frac{\|\mathbf{g} - \mathbf{R}\hat{\gamma}\|_2^2}{N - \text{sum}(w_i^{-1} \mathbf{C}_{\gamma\gamma})}, \quad (16)$$

where $\mathbf{C}_{\gamma\gamma}$ is the covariance matrix of the MAP at each iteration. As we can see, the denominator of the right hand in the equation (15) is negatively proportional to the diagonal of $\mathbf{C}_{\gamma\gamma}$, thus a small variance in $\mathbf{C}_{\gamma\gamma}$ leads to a large denominator, which contributes to eliminating the corresponding hyperparameter. Therefore, most of the hyperparameters will be driven to zero upon convergence and (15) amplifies the value at position of signals. In practice, the hyperparameters will not be driven to zero strictly but to some small values. We can achieve a sparse vector of hyperparameters by pruning those elements with small magnitude. Consequently, we can screen out the columns of steering matrix that are associated with the nonzero hyperparameters, hence reduce the system model. At each iteration, the estimate of reflectivity vector $\hat{\gamma}$ is obtained via conventional Tikhonov regularization with the estimate of $\hat{\mathbf{w}}$ and $\hat{\sigma}^2$ from the previous iteration. Such combination of Tikhonov regularization and reduction of system model contributes to a computationally efficient algorithm of TomoSAR inversion.

## 5 Experiments

### 5.1 Performance Evaluation on Angular Bias

To evaluate the performance of SBL with respect to PCA and KPCA approaches in TomoSAR, we implemented an experiment on a set of simulated data. We simulated 1000 samples. Each of them has a layover of 2 scatterers of random elevations that uniformly distributed between 0m and 300m. The amplitude ratio of the two scatterers were set to 1:2. We simulated SAR measurements on 13 baselines similar to that of TerraSAR-X repeat pass baselines. The baselines were distributed in the range $-200m$ to $200m$, thus leading to about 30m Rayleigh resolution unit. Noise were not included in the simulation, in order to test the ideal performance of the three algorithms. For comparison, we used the angular bias (i.e. inner product) between the estimated steering vectors and the true steering vectors as the criterion. The expression of the angular bias is shown as follows.

$$b = \cos^{-1}(\hat{\mathbf{r}}^H \mathbf{r}), \quad (17)$$

where $\hat{\mathbf{r}}$ is the steering vector estimate. The angular bias is always greater than 0, due to the definition of the arccosine function. For comparison, we calculated the mean and standard deviation of the angular bias, which were listed in Table 1 and 2.

|  | PCA | KPCA | SBL |
|---|---|---|---|
| First steering vector | | | |
| Mean | 18.0° | 7.4° | **1.5°** |
| Std | 17.0° | 14.0° | **7.2°** |
| Second steering vector | | | |
| Mean | 3.1° | 2.0° | **1.1°** |
| Std | 1.6° | 2.9° | **4.6°** |

**Table 1**. Mean and standard deviation of angular bias of the estimates of the first and the second steering vectors.

Additionally, we computed the percentage of both steering vectors estimates whose angular bias were less than 1°, 3°, 6°, and larger than 6°, respectively, in order to have a more intuitive view of the performance of the proposed algorithm.

| | PCA | KPCA | SBL |
|---|---|---|---|
| First steering vector | | | |
| <=1° | 0.8% | 0.66% | **61.9%** |
| <=3° | 5.0% | 50.5% | **97.9%** |
| <=6° | 13.5% | 88.0% | **98.5%** |
| >6° | 86.5% | 12.0% | **1.5%** |
| Second steering vector | | | |
| <=1° | 8.8% | 30.0% | **64.6%** |
| <=3° | 52.2% | 86.9% | **98.7%** |
| <=6° | 95.1% | 96.5% | **99.1%** |
| >6° | 4.9% | 3.5% | **0.9%** |

**Table 2**. Percentage of the first and the second steering vector estimates whose angular bias is less than 1°, 3°, 6°, and larger than 6°, respectively

From the tables and figure, we can see that the steering vectors estimates derived via SBL have much lower angular bias from our simulated data in general. It outperforms the other 2 methods by a factors of 2 to 12 in the mean angular bias. For more than 60% of the samples, the steering vectors can be estimated with less than 1° angular bias, whereas this number is only less than 1% for PCA- or KPCA-based methods. This gives a preliminary demonstration of the feasibility of SBL approach for high precision SAR tomographic inversion.

### 5.2 Demonstration of Prior Learning via SBL

We visualize the procedure of updating hyperparameters for a more intuitive view as follows. We simulated two individual scatterers layover in the same resolution unit. The two scatterers have the same phase and amplitude. The distance between the two scatterers was set to 0.6 Rayleigh resolution. In order to avoid the off-grid bias, we set the elevation of both scatters to be on-grid (1m grid in our simulation).. Noise was not includedd. Figure 1.a demonstrates the peaky signal reflected by the scatterers. figure 1.b to the figure 1.f show the learning procedure of the parameterized prior. We can find that the elevation position of the layovered scatterers can be determined once upon convergence. The position of the peak of the estimated parameterized prior correspond to the elevation position of the layovered scatterers.

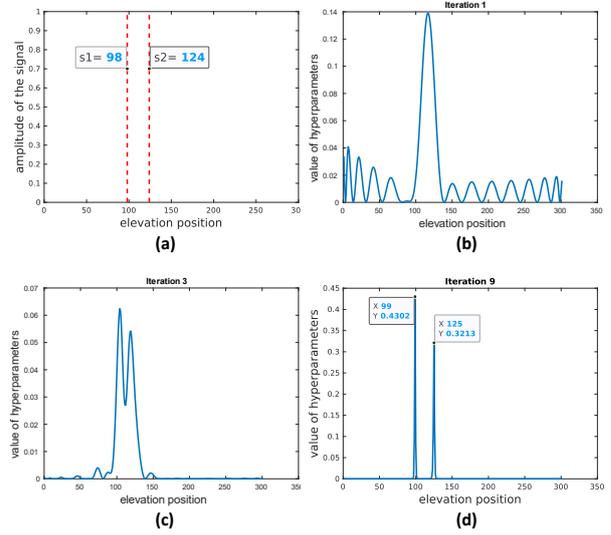

**Figure 1**. Visualization of learning procedure of the parameterized prior. (a), the amplitude and the true elevastion position of the two layovered scatterers. (b)-(d), the estimated hyperparameters, i.e. the diagonal of the covariance matrix $\mathbf{C}_{\gamma\gamma}$, during learning procedure.

### 5.3 Preliminary Study on the Super-resolution Power of SBL

We performed some experiments to inspect whether SBL is able to separate two closely spaced scatterers layovered in the same resolution unit. The experiments were carried out on a set of simulated data with similar simulation setting as [9]. Specifically, a stack containing 25 images was simulated, whose baselines were regularly distributed in the range -135m to 135m, thus leading to about 42m Rayleigh resolution. The noise was considered in this study and simulated with an SNR of 6dB. our simulation was performed with an increasing distance between the two scatterers, which is from 0.05 Reyleigh resolution till 1.25 Rayleigh resolution. The elevation of the first scatterer was simulated to follow an uniform distribution in the range 0m to 200m. We simulate the worst case where the amplitude and phase of the two scatterers are identical. For each set of data, 1000 samples were simulated. To evaluate the super-resolution power, the detection rate is an important indicator. We defined a succeccful detection as follows: (1) two scatterers were detected, (2) the position of two detected scatterers falled within 4 times Cramér–Rao bound (CRLB) with respect to the true elevation position.

figure 2 demonstrates the detection rate versus the normalized distance, which is the distance between the two scatterers, normalized to the Rayleigh resolution.

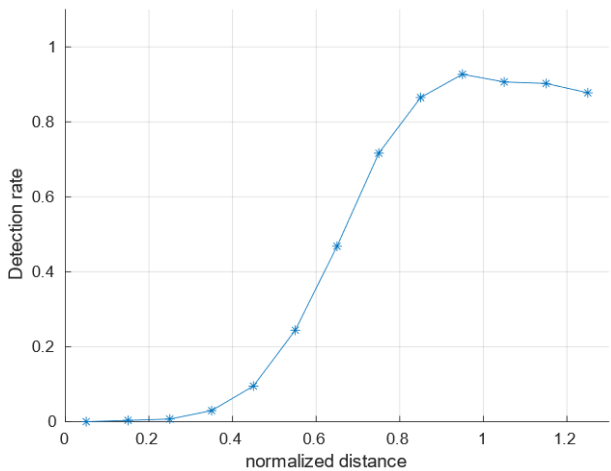

**Figure 2**. The detection rate versus the normalized distance at 6dB SNR.

As figure 2 shows, SBL achieves moderate super-resolution power in noisy cases. At around 0.7 Rayleigh resoltuion, SBL reaches about 60% detection rate. In addition, more than 80% double scatterers within the same resolution unit can be sucessesfully detected when the distance between double scatterers is larger than 0.8 Rayleigh resolution unit. Comparing to SVD-Wiener in [10][14], which is not able to overcome Rayleigh resolution and has little super-resolution power, SBL is obviously superior in the super-resolution capability at similar computational cost. Comparing to SL1MMER proposed in [9], SBL takes the advantage of high computational efficiency since it is fully based on iterative Tikhonov MAP and no compuational expensive optimization is required. Hence, SBL shows a great potential in large scale TomoSAR application in urban areas. In [10], we performed a systematical investigation about the super-resolution power of SBL.

## 6  Conclusion

This paper provides a data-driven approach for SAR tomographic inversion via sparse Bayesian learning. In the state-of-the-art compressive sensing based methods, such as "SL1MMER" [9], the sparsity of the estimated reflectivity profile is promoted by minimizing a predefined $L_1$ norm, thus approximating a $L_0$ norm minimization. However, the $L_1$ norm minimization requires a computationlally expensive optimization and hinders the application of compressive sensing based methods for large scale processing. SBL overcomes the high computational cost via iteratively estimating each element of the parameterized prior, with which we can solve the inversion problem through pure Tikhonov regularization and promote the sparsity of the estimated reflectivity profile at the meanwhile. Our experiments on simulated data show that SBL outperforms the PCA and KPCA-based method by a factor of 2 to 10 and 2 to 5, respectively, in terms of angular bias of the steering vectors estimates in noise-free cases. Comparing to PCA and KPCA, SBL takes both advantages of data-driven and model-driven methods. It solves the TomoSAR inversion problem with high computational efficiency and accuracy.

Moreover, a preliminary study if its super-resolution power demonstrates that the proposed SBL based TomoSAR iversion method can reach a moderate super-resolution capability under typical SNR cases like 6dB when reasonable number of images are available. This significantly reduce the computational cost comparing to the compressive sensing based approaches.

## 7  Literature

[1] G. Fornaro, S. Verde, D. Reale, and A. Pauciullo, "CAESAR: An Approach Based on Covari- ance Matrix Decomposition to Improve Multibaseline- Multitemporal Interferometric SAR Processing," *IEEE Trans. Geosci. Remote Sens.*, vol. 53, no. 4, pp. 2050–2065, Apr. 2015.

[2] Y. Wang and X. Zhu, "Robust Nonlinear Blind SAR Tomography in Urban Areas," in *EUSAR 2018, Aachen, Germany*, 2018.

[3] D. P. Wipf and B. D. Rao, "Sparse Bayesian learning for basis selection," *IEEE Trans. Signal Pro- cess.*, vol. 52, no. 8, pp. 2153–2164, Aug. 2004.

[4] X. Zhu and R. Bamler, "Tomographic SAR In-version by L1-Norm Regularization -- The Compres- sive Sensing Approach," *IEEE Trans. Geosci. Remote Sens.*, vol. 48, no. 10, pp. 3839–3846, 2010.

[5] Abramowitz, M.: Handbook of mathematical functions, 3rd ed., New York: Dover, 1980

[6] Guidelines for ETEP Authors, ETEP European Transactions on Electrical Power. Vol. 7, No. 5, Sept./Oct. 1997, pp. 363-364

[7] M. Tipping, "Sparse bayesian learning and the relevance vector machine," *J. Mach. Learn. Res.*, Sep. 2001.

[8] D. J. C. MacKay, "Bayesian interpolation," *Neural Comput.*, vol. 4, no.3, pp. 415–447, 1992.

[9] X. X. Zhu and R. Bamler, "Super-Resolution Power and Robustness of Compressive Sensing for Spectral Estimation With Application to Spaceborne Tomographic SAR," in *IEEE Transactions on Geoscience and Remote Sensing*, vol. 50, no. 1, pp. 247-258, Jan. 2012.

[10] Y. Wang, K. Qian and X.X Zhu, "On Super-resolution Power of SAR Tomographic Inversion via Sparse Bayesian Learning", *under submission*.

[11] Z. Yang, L. Xie and C. Zhang, "Off-grid Direction of Arrival Estimation Using Sparse Bayesian Inference", arXiv:1108.5838v4.

[12] J. Wang and Rui Min, "An approach for sparse baselines SAR tomography," *Proceedings of 2011 IEEE*

*CIE International Conference on Radar*, Chengdu, 2011, pp. 1452-1455.

[13] R. Min, Y. Hu, Y. Pi, Z. Cao, "SAR Tomography Imaging Using Sparse Bayesian Learning", IEICE Transactions on Communications, 2012, Volume E95.B, Issue 1, Pages 354-357, Released January 01, 2012.

[14] X. Zhu, R. Bamler, "Very high resolution spaceborne SAR tomography in urban environment", IEEE Transactions on Geoscience and Remote Sensing, vol. 48, issue. 12, 2010, pp. 4296-4308.